\documentstyle[12pt]{article}
\oddsidemargin--1mm
\begin{document}
\newcommand{\pl}{\partial}
\newcommand{\be}{\begin{equation}}
\newcommand{\ee}{\end{equation}}
\newcommand{\ba}{\begin{eqnarray}}
\newcommand{\ea}{\end{eqnarray}}
\newcommand{\mbf}[1]{\mbox{\boldmath$ #1$}}
\def\<{\langle}
\def\>{\rangle}

\noindent
\hspace*{10cm} {\small\sf IFT preprint UFIFT-MATH-02-1}  

\begin{center}
{\Large\bf On a low energy bound in a class of chiral field theories
with solitons}

\vskip 0.5cm
Sergei V. Shabanov {\footnote{on leave from Laboratory of
Theoretical Physics, JINR, Dubna, Russia}}

\vskip 0.2cm

{\em Institute for Fundamental Theory, 
Departments of Physics and Mathematics,\\ University of Florida,
Gainesville, FL- 32611, USA}
\end{center}

\begin{abstract}
A low energy bound in a class of chiral solitonic field theories
related the infrared physics of the SU(N) Yang-Mills theory is established. 
\end{abstract}

{\bf 1}. {\bf The model}. 
Consider $N-1$ smooth fields $n_a=n_a(x)$ in spacetime taking their values 
in the Lie algebra of $SU(N)$. The fields are chosen to be commutative
$[n_a,n_b]=0$ and orthonormal $(n_a,n_b)=\delta_{ab}$ with respect the 
Cartan-Killing form in the Lie algebra. For any two Lie algebra elements
$\xi$ and $\eta$, the Cartan-Killing form is defined
as $(\xi,\eta) = {\rm tr}\, (\hat{\xi}\hat{\eta})$ where the operator
$\hat{\xi}$ acts on the Lie algebra as a Lie derivative $\hat{\xi}\eta=
[\xi,\eta]$. There can only be $N-1$
mutually commutative and linearly independent elements in the Lie algebra
of $SU(N)$ because the rank of $SU(N)$ is $r=N-1$ (the dimension of the 
Cartan subalgebra). If $h_a$ form an orthonormal basis in the Cartan
subalgebra in a matrix representation of SU(N), then 
\be
n_a(x) = U^\dagger(x) h_aU(x)\ ,
\label{1}
\ee 
where $U(x)\in SU(N)$.
  In Eq. (\ref{1}) $U(x)$ is defined
modulo the left multiplication by elements 
from  the Cartan subgroup generated by $h_a$ 
(the maximal Abelian subgroup $T= U(1)^{N-1}$).
So, in fact, $U(x) \in SU(N)/T$ 
since any group element can be represented as a product of an element
of $T$ and an element of the quotient $SU(N)/T$. Under the condition that
$n_a$ approach fixed constant values at the spatial infinity,
$n_a(x)\rightarrow h_a$, i.e., $U(x)$ approaches the group unity,
 the fields $n_a$ define a map of a spatial
three-sphere ${\bf S}^3$ into the manifold
$SU(N)/T$ for every moment of time. 
The third homotopy group of this map is nontrivial 
$\pi_3(G/T)\sim Z$, $G=SU(N)$.   
When $N=2$, the only field $n_1$ can be regarded as a unit 3-vector.
It is a Hopf map: ${\bf S}^3
\rightarrow {\bf S}^2 \sim SU(2)/U(1)$. The corresponding topological
number is the Hopf invariant which can also be interpreted as a linking
number of two curves in ${\bf S}^3$ being preimages of two distinct
points of ${\bf S}^2$. The two-forms $F^a = F^a_{jk}dx^j\wedge dx^k$,
$j,k=1,2,3$, where
\be
F^a_{jk} = iN{\textstyle{\sum_b}}(n_a, [\pl_jn_b,\pl_kn_b])\ ,
\label{2}
\ee
are closed, that is, $F^a_{jk} = \pl_jC^a_k -\pl_kC^a_j$.
This is proved at the end of next section. The forms $F^a$ may not
be exact. This follows from the fact that the cohomology ring $H^*(G/T)$
is rationally generated by $H^2(G/T)$ \cite{bott}. The topological number of 
the map ${\bf S}^3\rightarrow G/T$ should be constructed out of the 2-forms 
$F^a = iN(n_a,\sum_b[dn_b,dn_b])$ on $G/T$. Introducing 
the field $B_i^a = \frac 12\epsilon_{ijk}F^a_{jk}$ with
$\epsilon_{ijk}$ being the Levi-Chevita tensor, the topological
number of the above map can be written as
\be
Q = (16\pi^2 N)^{-1}\,\int dx\, {\textstyle{\sum_{i,a}}}C^a_iB^a_i\ .
\label{3}
\ee
For SU(2), $Q$ is a Hopf invariant. Since $SU(2)/U(1)\subset G/T$,
the normalization 
coefficient in (\ref{3}) can be chosen so that $Q$ is an integer when
$n_a$ realize a Hopf map. 

An explicit realization of the Hopf map by
the fields $n_a$ is as follows. 
Consider the Cartan-Weyl basis in the Lie algebra.
Let $\alpha$ be a positive root. For every positive root $\alpha$, there
are two  basis elements $e_{\alpha}$ and $e_{-\alpha}=\bar{e}_{\alpha}$ 
such that for any element $h$
from the Cartan subalgebra     
\ba\label{4}
[h, e_\alpha ]&=&(h,\alpha)\, e_\alpha\ ,   \\ \label{5}
[{e}_{\alpha},{e}_{-\alpha}] &=& \alpha\ ,  \ \ \ \ \    
[{e}_\alpha , {e}_\beta] = N_{\alpha,\beta}\, e_{\alpha+\beta} \ ,
\ea
where $N_{\alpha,\beta}\neq 0$ if  $ \alpha+\beta$ is a root.
Note that the elements $\alpha,\ e_\alpha$ and $e_{-\alpha}$ form a basis
of an SU(2) subalgebra (associated with the root $\alpha$).
Let $U_\omega(x)\in SU(2)/U(1) \subset SU(N)/T$ where the subgroup SU(2)
is associated with a simple root $\omega$. One can always choose
$h_1 = N^{1/2} \omega$. The norm of any root of SU(N) is $1/N$ with respect
to the Cartan-Killing form (see next section).  
Then $n_1(x) =N^{1/2} U^\dagger_\omega \omega U_\omega$ 
is a Hopf map. 
The other fields realize a trivial map, $n_a =h_a, \ a>1$.
Indeed, $U_\omega (x) = \exp[{iu_\omega(x)}]$ where $u_\omega(x)=
\varphi_{\omega}(x)e_\omega + \bar{\varphi}_\omega(x) e_{-\omega}$.
For $a>1$, it follows from (\ref{4}) that 
$n_a = U_\omega^\dagger h_a U_\omega = h_a$ because
$(\omega, h_a)\sim (h_1,h_a) =0$. Now, if we introduce an orthonormal
basis in the SU(2) subgroup, $\tau_1 =i(e_{\omega} -e_{-\omega})/\sqrt{2}$,
$\tau_2 =(e_{\omega} +e_{-\omega})/\sqrt{2}$ and $\tau_3=\sqrt{N}\omega$,
then $[\tau_j,\tau_k]=iN^{-1/2}\epsilon_{jkn}\tau_n$. Let $\mbf{n}$ be
an isotopic unit three-vector whose components are $(\tau_j,n_1)$. It defines
the Hopf map by construction. From (\ref{2}) we infer that 
$F_{jk}^a = \delta^{a1}\sqrt{N} \mbf{n}\cdot(\pl_j\mbf{n}\times\pl_k\mbf{n})$.
Hence our $B_j^a$ and $C_j^a$ contain an extra factor $\sqrt{N}$ when
the fields $n_a$ realize a Hopf
map associated with an SU(2) subgroup of SU(N). This explains the normalization 
factor $N^{-1}$ in (\ref{3}). 
Since all the root have the same norm in SU(N),
the normalization coefficient in (\ref{3}) for any SU(2) subgroup
has to be the same. The root system is invariant under the Weyl symmetry,
and so should be $Q$. The sum over $a$ in (\ref{3}) provides this invariance.

The dynamics of the fields $n_a$ is determined by the Lagrangian density
\be\label{7}  
{\cal L} = m^2 \sum_{\mu, a}(\pl_\mu n_a,\pl_\mu n_a) - \frac{g }{4} 
\sum_{\mu,\nu,a} F_{\mu\nu}^aF_{\mu\nu}^a\ ,  
\ee
$\mu,\nu = 0,1,2,3$; and $\pl_0$ stands for the time derivative. 
In the case of SU(2), this Lagrangian density describes the 
Faddeev model \cite{f} for 
knot solitons. The knot solitons have been extensively studied
numerically \cite{knots}.
The model (\ref{7}) has been introduced in \cite{fn1} and may also
have solitonic solutions. The Lagrangian (\ref{7}) 
is believed to describe (in a certain
approximation) the infrared physics of the SU(N) Yang-Mills theory 
\cite{fn1,fn2,sh1}. Recent analytical \cite{gies} and lattice \cite{sh2,jena}
studies of this correspondence in the SU(2) case look promising. 

Due to the Lorentz symmetry of the Langrangian
density, a Lorentz transformation
of a static solution is a time dependent solution of the Euler-Lagrange
equations for  (\ref{7}). Solutions that describe interacting solitons
are not static (even modulo Lorentz transformations). 
In this paper a low energy bound for static solitons with a topological
number $Q$ is established: 
\ba\label{8}
E &\geq& c_N |Q|^{3/4}\ ,\\ 
\label{cn}
c_N&=& 8\pi^2 3^{3/8}\, \left(\frac{2N^3}{N^2-N-1}\right)^{1/4}\sqrt{m^2g}\ ,
\ea 
where $E$ is the energy functional
\ba\label{e1}
E&=& m^2 \int dx \sum_{j,a}(\pl_j n_a, \pl_j n_a) + \frac{g }{2}\int dx 
\sum_{j,a} B_j^aB_j^a  \\
\label{e3}
&\equiv & \int dx\left( {\cal E}_2(x) +{\cal E}_4(x)\right)
\equiv E_2 +E_4\ . 
\ea
For the Faddeev-Hopf knot solitons the low energy bound was found
in \cite{bound1} and improved in \cite{bound2,bound3} (meaning a larger
constant $c_2$). Beyond conventional perturbation theory, 
the Yang-Mills quantum dynamics can be studied by 
the large $N$ expansion method with the purpose to establish
a relation (duality) to a string theory on some manifold.  
Therefore it is of interest to investigate
the $N$ dependence of the low energy bound for solitons in the model
(\ref{7}).     

{\bf 2. Notations and necessary facts}. 
We would need the following algebraic inequalities. Let
$ a_i,\, b_i\geq 0$, $a=\sum_i a_i$, $b=\sum_i b_i$ and  $\gamma \geq 1$. Then
\ba\label{2.6}
a_1^\gamma + a_2^\gamma +\cdots +a_r^\gamma     &\leq& 
a^\gamma\ ;\\ 
\label{p6} \sqrt{a}\leq 
 \sqrt{a_1} + \sqrt{a_2}+\cdots +\sqrt{a_r}&\leq& \sqrt{r}\sqrt{a}\ ;\\
\label{new1}
{\textstyle{\sum_i}} a_i^pb_i^q &\leq& a^pb^q\ ,\ \ \ \ p+q=1\ .
\ea
Define $p_i = a_i/a \leq 1$. Then $\sum_i p_i =1$.
The inequality (\ref{2.6}) follows from an obvious inequality
$p_i^\gamma \leq p_i$ if one takes the sum over $i$.
The second inequality is proved by squaring it and applying 
the basic algebraic inequality
$\sqrt{a_i}\sqrt{a_j}\leq \frac 12(a_i +a_j)$. The third inequality
is an algebraic H\"older inequality (see, e.g., \cite{klauder}).  

An arrow is used to denote
vectors in space, e.g., $\vec{\pl} \phi =(\pl_1 \phi,\pl_2\phi,\pl_3 \phi)$
for the gradient. The scalar product for two vector fields is
\be \nonumber
\< \vec{u},\vec{v}\,\> =\int dx\, \vec{u}\cdot\vec{v}\ .
\ee
The $L_p$ norm of a vector field reads
\be
\Vert \vec{u}\Vert_p = \left[\int dx \left(\vec{u}\cdot\vec{u}\,\right)^{p/2}
\right]^{1/p}\label{2.2}\ .
\ee     
The following functional inequalities are used in the sequel
\ba\label{2.3}
\vert\<\vec{u},\vec{v}\,\>\vert &\leq& \Vert\vec{u}\Vert_p\Vert\vec{v}
\Vert_{q}\ , \ \ \ \ p^{-1} +{q}^{-1} =1\ , \\ \label{2.4}
\Vert\vec{u}\,\Vert_{6/5}&\leq& \Vert\vec{u}\,\Vert^{2/3}_1\,
\Vert\vec{u}\,\Vert^{1/3}_2\ , \\ \label{2.5}
\Vert\vec{u}\,\Vert_6 &\leq& \lambda_1\Vert{\sf curl}\,\vec{u}\,\Vert_2
\ ,\ \ \ \  \lambda_1 = (48)^{1/6}(3\pi)^{-2/3}\ .
\ea
The first two inequalities are H\"older type inequalities \cite{klauder}. 
The third one follows from Rosen's result for
scalar functions \cite{rosen} (cf. also \cite{bound2})
\be
\label{rosen1}
\Vert\phi\Vert_6 \leq \lambda_1\,\Vert\vec{\pl}\phi\Vert_2\ ,
\ee
where the $L_p$ norm for scalar functions is defined by (\ref{2.2})
for one-dimensional vectors. Let $\phi = (\vec{u}\cdot\vec{u})^{1/2}$.
We have
$$
\vec{\pl}\phi\cdot\vec{\pl}{\phi} = \phi^{-2}{\textstyle{\sum_j}}
(\pl_j\vec{u}\cdot\vec{u})^2 \leq 
{\textstyle{\sum_j}}\pl_j\vec{u}\cdot\pl_j\vec{u}\ .
$$
Making use of this inequality, we infer (\ref{2.5}) from (\ref{rosen1}):
\be\label{rosen2}
\Vert \vec{u}\Vert_6 = \Vert \phi\Vert_6
\leq \lambda_1\,\Vert\vec{\pl}\phi\Vert_2
 \leq\lambda_1 \left[\int dx 
{\textstyle{\sum_{i,j}}} (\pl_j u_i)^2\right]^{1/2}= \lambda_1
\Vert{\sf curl}\,\vec{u}\,\Vert_2\ . 
\ee
The last equality in (\ref{rosen2}) is true if
${\sf div}\, \vec{u}=0$ and $\vec{u}$ decreases sufficiently fast at spatial
infinity, which we require in  (\ref{2.5}). The coefficient $\lambda_1$ is
the least possible coefficient in inequality (\ref{2.5}) \cite{rosen}.  

Let $\omega_a$, $a=1,2,..., r$, be simple roots of SU(N). 
They have the same norm $(\omega_a,\omega_a) \equiv \gamma^2$.
The angle between $\omega_{a}$ and $\omega_{a\pm 1}$ is $2\pi/3$, and
otherwise the roots are perpendicular.
Any positive root can be written in the form $\alpha = \omega_a + \omega_{a+1}
+\cdots +\omega_{a+q}$ for $a+q\leq r$.  From this it is easy to deduce
that all roots have the same norm with respect to the Kartan-Killing form,
$(\alpha, \alpha)=\gamma^2$. To find the actual norm $\gamma$,
one should compute, say, the matrix $\hat{\omega}_1$ in the Cartan-Weyl basis
and take the trace of its square. 
From (\ref{4}) it follows that $\hat{\omega}_1$ is block diagonal.
The block associated with the Cartan subalgebra is zero because $\omega_a$
commute amongst each other. The nontrivial blocks come from the subspaces
spanned by $e_\alpha$ and $e_{-\alpha}$ where the positive root $\alpha$
is either equal to $\omega_1$ or contains $\omega_2$ or $\omega_1+\omega_2$.  
There are $r-1$ roots containing $\omega_2$ and $r-1$ roots containing
$\omega_1+\omega_2$. Then
$\gamma^2 = {\rm tr}\,(\hat{\omega}^2_1) = \gamma^4 N$ as
is deduced from (\ref{4}). Hence 
\be
(\omega_a,\omega_a) = N^{-1}\ .
\label{2.7}
\ee
As a consequence of (\ref{2.7}), the following identity holds for any
Lie algebra element $v$
\be
v = \sum_a n_a (n_a,v) + N \sum_a [n_a,[n_a,v]]\equiv v_\Vert +v_\perp\ .
\label{2.8}
 \ee
The proof is based on the following observation. Relation (\ref{2.8})
is covariant under the adjoint action of SU(N). So, according to (\ref{1}),
$n_a$ can be replaced by $h_a$ after a corresponding adjoint rotation of
$v$. Decomposing $v$ in the Cartan-Weyl basis, one can see that the 
first term in (\ref{2.8}) is the Cartan component of $v$. The double
commutator in the second term can be computed by means of (\ref{4}) and gives
rise to
the factor $\sum_a (\alpha,h_a)^2 = (\alpha,\alpha) =N^{-1}$ for every
basis element $e_{\alpha}$. Thus the second term in (\ref{2.8}) is nothing
but a projector onto the subspace orthogonal to the Cartan subalgebra spanned
by $n_a$.

By differentiating (\ref{1}) one finds 
\ba\label{2.9}
\pl_\mu n_a +i[A_\mu,n_a]=0\ ,  \\ \label{2.9a}
i\pl_\mu U^\dagger U \equiv A_\mu - \sum_a n_a C_\mu^a\ ,\ \ \ (n_a,A_\mu)=0
\ .
\ea
Equation (\ref{2.9}) can be interpreted as: The fields $n_a$ are transported
parallel with respect to the connection $A_\mu$. Taking a commutator 
of (\ref{2.9}) with $n_a$, summing over $a$ and making use of the identity
(\ref{2.8}), the connection $A_\mu$ can be explicitly written via $n_a$,
\be
A_\mu = iN\sum_a [\pl_\mu n_a, n_a] \ . \label{a}    
\ee
The connection (\ref{a}) has been introduced by Cho to study 
monopoles in the Yang-Mills theory for SU(2) and SU(3) \cite{cho}.
By multiplying (\ref{2.9}) by $n_b$ using the Cartan-Killing form, 
one deduces that the derivatives of $n_a$ are orthogonal
to the fields themselves
\be
(\pl_\mu n_a,n_b)=0 \ .
\label{2.8b}
\ee

Now we show that the tensor (\ref{2}) is an Abelian gauge field tensor
(cf. \cite{fn1,sh1}),
that is, the two-forms $ F^a$ are closed, $dF^a =0$. Consider the following
algebraic transformations
\ba\label{i2}
\pl_\mu A_\nu - \pl_\nu A_\mu &=& 2iN{\textstyle{\sum_a}}[\pl_\nu n_a,\pl_\mu
n_a] = 2iN{\textstyle{\sum_a}}\left[
[A_\mu,n_a],[A_\nu,n_a]
\right] \\ \label{i3}
&=&2iN{\textstyle{\sum_a}}\left\{
[A_\nu, [n_a,[n_a,A_\mu]]] +[n_a,[[n_a,A_\mu],A_\nu]]
\right\} \\ \label{i4}
&=& 2i[A_\nu,A_\mu] +iN{\textstyle{\sum_a}}[n_a,[n_a,[A_\mu,A_\nu]]]\\
\label{i5}
&=& -i[A_\mu,A_\nu] - i{\textstyle{\sum_a}}n_a\,(n_a,[A_\mu,A_\nu])\ .
\ea 
In (\ref{i2}) we have used (\ref{a}); next, $\pl_\mu n_a$ has been transformed
via (\ref{2.9}); 
Eq. (\ref{i3}) follows from the Jacobi identity; to derive (\ref{i4}), the
first term in (\ref{i3}) has been transformed by means of the algebraic 
identity (\ref{2.8}), while the second one via the Jacobi identity; finally,
by applying the algebraic identity (\ref{2.8}) to (\ref{i4}), Eq. (\ref{i5}) 
has been deduced. Introducing the Yang-Mills field strength tensor
\be
F_{\mu\nu} = \pl_\mu A_\nu -\pl_\nu A_\mu +i[A_\mu,A_\nu]\ ,
\ee
it follows from (\ref{i5}) that
\be\label{i6}
F_{\mu\nu} = i{\textstyle{\sum_a}}n_a\,(n_a,[A_\mu,A_\nu]) =
{\textstyle{\sum_a}}n_a F_{\mu\nu}^a\ .
\ee
The last equality in (\ref{i6}) is deduced 
by multiplying (\ref{i5}) and the middle of (\ref{i2})
by $n_a$ using the Cartan-Killing form. Now observe that the field
strength (curvature) 
of the pure gauge connection (\ref{2.9a}) is zero. Making use of
the decomposition (\ref{2.9a}) of a pure gauge connection we obtain
\be
0=F_{\mu\nu} -  {\textstyle{\sum_a}}n_a \left(\pl_\mu C_\nu^a -\pl_\nu C_\mu^a
\right) \ ,\nonumber
\ee
where the identity (\ref{2.9}) has been used again 
for algebraic transformations. Thus, 
$F_{\mu\nu}^a =\pl_\mu C_\nu^a -\pl_\nu C_\mu^a$. Note that (\ref{2.9a})
allows one to determine $C_\mu^a$ via the group element $U$ explicitly.
In (\ref{3}) the vector potential $C_i^a$ can always be chosen to satisfy
the Coulomb gauge, $\pl_iC_i^a=0$, thanks to the gauge freedom 
$C_\mu^a\rightarrow C_\mu^a +\pl_\mu \xi^a$.

{\bf 3. A key algebraic inequality}. In this section the following inequality
is proved
\be\label{s1}
{\cal E}_4 \leq \kappa_N{\cal E}_2^2\ ,\ \ \ \ \kappa_N= \frac {gN}{4m^4}\left(
1-\frac{1}{N^2-N}\right)\ .
\ee
It is used in the next section to establish the low energy bound.
Consider the $(N^2-1)\times(N^2-1)$ matrix
$$
G = \sum_{i,b}\pl_in_b\otimes\pl_in_b\ .
$$
It can be regarded as a linear operator on the Lie algebra, i.e.,
$G\eta = \sum_{i,b}\pl_in_b (\pl_in_b,\eta)$ for any Lie algebra
element $\eta$. It has $N-1$ zero eigenvalues because $Gn_a=0$.
The matrix $G$ satisfies  
\be \label{s3}
{\rm tr}\, G^2 \geq \frac{1}{N^2-N}\left({\rm tr}\,G\right)^2\ .
\ee 
The proof is simple. Let $g_k$, $k=1,2,..., n=N^2-N$, 
be nonzero eigenvalues of $G$. They are real since $G=G^T$ with
respect to the Cartan-Killing form.
Consider a function of one real variable $\xi$, 
$f(\xi) = \sum_k (g_k - \xi)^2 $. Computing the sum explicitly,
one finds that $f(\xi) = {\rm tr}\, G^2 -2\xi{\rm tr}\,G + \xi^2n$.
The function attains its absolute minimum at $\xi =\xi_0 = {\rm tr}\,G/n$.
Since $f(\xi)\geq 0$ for all $\xi$'s, the inequality (\ref{s3}) follows
from $f(\xi_0)\geq 0$.

Consider a local Cartan-Weyl basis which is obtained by an 
adjoint transformation of the basis (\ref{4}), (\ref{5}) 
with the group element $U(x)$ defined in (\ref{1}). 
Denoting $n_\alpha = U^\dagger e_\alpha U$ and
$n_{-\alpha} = U^\dagger e_{-\alpha} U=\bar{n}_\alpha$ we have
\be\label{lw}
[n_a,n_\beta] = (h_a,\beta) n_\beta\ ,\ \ \ \ 
[n_\alpha,n_{-\alpha}] =U^\dagger \alpha U\equiv \alpha_U\ ,
\ee
where $(\alpha_U,n_a)=(\alpha,h_a)$ and $(n_a,n_\beta)=0$,
$(n_\alpha,n_{-\alpha})=1$,
$(n_\alpha,n_{\alpha})=(n_{-\alpha},n_{-\alpha})=0$.

To establish a relation between ${\cal E}_4$, ${\cal E}_2$ and $G$,
we decompose the connection (\ref{a}) in the local
Cartan-Weyl basis
\be\label{s4}
A_i = \sum_{\alpha >0}\left(A_i^\alpha n_\alpha  + c.c.\right)
\equiv H_i + \bar{H}_i \ .
\ee
Then we obtain 
\be\label{s5}  
{\cal E}_2= m^2{\rm tr}\,G =-m^2\sum_{i,b}([A_i,n_b],[A_i,n_b]) =
\frac {m^2}{N} \sum_i (A_i,A_i) = \frac {2m^2}{N} \sum_i (\bar{H}_i,H_i)\ .
\ee
The second equality follows from (\ref{2.9}); the third one is 
a consequence of (\ref{2.8}) and (\ref{2.9a}).
Making use of (\ref{i6}) we also get
\ba\label{s6}
{\cal E}_4 &=& -\frac g4\sum_{a,i,j}\left(n_a,[A_i,A_j]\right)^2\\
\nonumber
&=& -\frac g4 \sum_{\alpha,\beta >0}\sum_{i,j}
\left( A_i^\alpha \bar{A}_j^\alpha - c.c.\right)(\alpha,\beta)
\left( A_i^\beta \bar{A}_j^\beta - c.c.\right)\\
\label{s8}
&\leq& -\frac{g}{4N}\sum_{i,j}\left[(H_i,\bar{H}_j) - c.c.\right]^2\ .
\ea
Note that the local Cartan component of $[A_i,A_j]$  can only come
from the second commutation relation in (\ref{lw}). Hence the sum over $a$
in (\ref{s6}) yields the factor $\sum_a(\alpha,h_a)(h_a,\beta)=
(\alpha,\beta) = N^{-1}\cos\theta_{\alpha\beta}\leq N^{-1}$ for
any two positive roots $\alpha$ and $\beta$ with the angle 
$\theta_{\alpha\beta}$ between them. 
In a similar fashion we derive
\ba\label{s9}
{\rm tr}\,G^2 &=& \sum_{a,b}\sum_{i,j}\left([A_i,n_a],[A_j,n_b]\right)^2
=\sum_{a,b}\sum_{i,j}\left(n_a,[[n_b,A_i],A_j]
\right)^2\\
\nonumber
&=&\sum_{\alpha,\beta >0}\sum_{i,j}
\left( A_i^\alpha \bar{A}_j^\alpha + c.c.\right)(\alpha,\beta)^2
\left( A_i^\beta \bar{A}_j^\beta + c.c.\right)\\
\label{s12}
&\leq& \frac{1}{N^2}\sum_{i,j}\left[(H_i,\bar{H}_j) + c.c.\right]^2\ .
\ea
Combining (\ref{s12}) and (\ref{s8}), we infer
\be\label{s14}
4Ng^{-1}{\cal E}_4 + N^2{\rm tr}\, G^2 \leq
4\sum_{i,j}(H_i,\bar{H}_j)(H_j,\bar{H}_i)\\
\leq 4\left(\sum_{i}(H_i,\bar{H}_i)\right)^2 = 
4\left({\rm tr}\,G\right)^2\ ,
\ee
where the Schwartz inequality
$$
\left\vert(H_i,\bar{H}_j)\right\vert^2 \leq (H_i,\bar{H}_i)(H_j,\bar{H}_j)
$$
has been used. 
The inequality (\ref{s1}) immediately follows from (\ref{s14}),
(\ref{s5}) and (\ref{s3}).

{\bf 4. The low energy bound}. Let $\lambda_0 = (16\pi^2 N)^{-1}$. Then
Eq. (\ref{3}) can be written as $Q= \lambda_0\sum_a\<\vec{C}^a,\vec{B}^a\>$.
Making use of (\ref{2.3}) one gets (cf. the case $N=2$ in \cite{bound1})
\ba\label{p1}
|Q| &\leq& \lambda_0\sum_a \Vert\vec{C}^a\Vert_p\Vert\vec{B}^a\Vert_{p^\prime}
\\ \label{p1a}
&\leq& \lambda_0\lambda_1\sum_a \Vert{\sf curl}\,
\vec{C}^a\Vert_2\Vert\vec{B}^a\Vert_{6/5}  \\
\nonumber
&=& \lambda_0\lambda_1 \sum_a\Vert\vec{B}^a\Vert_{2}
\Vert\vec{B}^a\Vert_{6/5}\\
\label{p2}
&\leq& \lambda_0\lambda_1 \sum_a\Vert\vec{B}^a\Vert_{2}
\Vert\vec{B}^a\Vert_{2}^{1/3}\Vert\vec{B}^a\Vert_{1}^{2/3}\ .
\ea
To get (\ref{p1a}),  Eq. (\ref{2.5}) has been used, which 
dictated the choice $p=6$ in (\ref{p1}), and also $\pl_iC^a_i=0$; 
then the substitution 
$\vec{B}^a ={\sf curl}\, \vec{C}^a$ has been made; the last inequality
(\ref{p2}) is a consequence of  (\ref{2.4}).
The energy can be written as 
\be
E=m^2\sum_a \Vert\vec{\pl}\,n_a\Vert_2^2 + \frac{g }{2}
\sum_a \Vert\vec{B}^a\Vert_2^2= \sum_a(E_{2a} +E_{4a})= E_2 +E_4\ .
\ee
Hence, continuing (\ref{p2}) we get
\ba\label{p4}
|Q|&\leq& \lambda_0\lambda_1 \left(2/g \right)^{2/3} 
\sum_a \left(E_{4a}^{4/3}\right)^{1/2}
\left(\Vert\vec{B}^a\Vert_{1}^{4/3}\right)^{1/2}\\
\label{p4a}
&\leq&
\lambda_0\lambda_1 \left(2/g\right)^{2/3}
\left[\left(\sum_a E_{4a}\right)^{4/3}\right]^{1/2}
\left[\left(\sum_b\Vert\vec{B}^b\Vert_{1}\right)^{4/3}\right]^{1/2}\\
\nonumber
&=&
\lambda_0\lambda_1 \left(2/g\right)^{2/3}
E_4^{2/3}\left[\sum_b\int dx \sqrt{\vec{B}_b\cdot\vec{B}_b}\right]^{2/3}\\
\label{p4c}
&\leq&\lambda_0\lambda_1 g^{-1}
\left(2E_4\right)^{2/3} (N-1)^{1/3}\left[\int dx 
\sqrt{2{\cal E}_4}\right]^{2/3}\\
\label{p4d}
&\leq&\lambda_0\lambda_1 g^{-1} \left(2 E_4E_2\right)^{2/3}
\left[2(N-1)\kappa_N\right]^{1/3}\\
\label{p4e}
&\leq&
\lambda_0\lambda_1 g^{-1}2^{-2/3}\left[2(N-1)\kappa_N\right]^{1/3}
E^{4/3}\ \\
\nonumber
&=&c_N^{-4/3} E^{4/3}\ ,
\ea
where the constant $c_N$ is given in (\ref{cn}).
To get (\ref{p4a}), the H\"older inequality (\ref{new1})
for $p=q=1/2$ and then (\ref{2.6}) for $\gamma =4/3$
have been applied; (\ref{p4c}) follows from (\ref{p6}); the 
algebraic inequality (\ref{s1}) has been used to deduce (\ref{p4d});
the final result (\ref{p4e}), which is equivalent to (\ref{8}),
comes from the basic inequality $E_4E_2 \leq E^2/4$.

 If the Lagrangian (\ref{7}) defines
an effective theory of the SU(N) gauge fields in some approximation, then
the coefficients $m$ and $g$  should depend on $N$, the Yang-Mills coupling
constant and a mass scale that determines the energy range in which 
the approximation is valid. 

The result (\ref{8}) is trivially generalized  to the case
when the mass scales and coupling constants are different for each
mode $n_a$, that is, $m^2$ and $g$ are replaced by $m_a^2$ and $g_a$, 
respectively, and inserted into the corresponding sums over $a$ in (\ref{7})
(cf. \cite{fn1}). In this case,  $m^2=\max_a \{m_a^2\}$ and
$g=\max_a\{g_a\}$ in (\ref{cn}). Indeed, all the inequalities in sections
3 and 4 still hold as a consequence of $m^2_a\leq m^2$ and $g_a\leq g$.
Equality (\ref{s5}) becomes an inequality $m^2{\rm tr}\,G\leq {\cal E}_2$,
which, however, does not affect our derivation of (\ref{s1}) from
(\ref{s14}).

{\bf Acknowledgments}. I wish to thank John Klauder and David Metzler
for useful discussions, David Groisser for the reference \cite{bott}, and
Richard Ward for helpful comments and the reference \cite{rosen}.

\end{document}